\documentstyle[12pt]{article}
\begin{document}

\newcommand{\nd}[1]{/\hspace{-0.6em} #1}
\begin{titlepage}
\begin{flushright}
CERN-TH.6351/91 \\
ACT-55\\
CTP-TAMU-100/91\\
\end{flushright}
\begin{centering}
\vspace{.1in}
{\large {\bf Quantum Mechanics and Black Holes in Four-Dimensional
String Theory }} \\

\vspace{.4in}
{\bf John Ellis} and {\bf N.E. Mavromatos}\\
\vspace{.05in}
Theory Division, CERN, CH-1211, Geneva 23, Switzerland  \\
and \\

\vspace{.05in}
{\bf D.V. Nanopoulos}\\

\vspace{.05in}
Center for Theoretical Physics, Dept. of Physics, \\
Texas A \& M University, College Station, TX 77843-4242, USA \\
\vspace{.03in}
and \\
\vspace{.05in}
Astroparticle Physics Group \\
Houston Advanced Research Center (HARC),\\
The Woodlands, TX 77381, USA\\
\vspace{.03in}
and \\
Theory Division, CERN, CH-1211, Geneva 23, Switzerland \\
\vspace{.1in}
{\bf Abstract} \\
\vspace{.05in}
\end{centering}
{\small
\paragraph{}
In
previous papers we have shown how strings
in a two-dimensional target space reconcile quantum
mechanics with general relativity, thanks to an infinite
set of conserved quantum numbers, ``W-hair'', associated with
topological soliton-like states. In this paper we extend
these arguments to four dimensions, by considering explicitly
the case of string black holes with radial symmetry.
The key infinite-dimensional W-symmetry is associated with the
$\frac{SU(1,1)}{U(1)}$ coset structure of the dilaton-graviton sector
that is a model-independent feature of spherically symmetric
four-dimensional
strings.
Arguments are also given
that
the enormous
number of string  {\it discrete (topological)}
states
account for the maintenance of
quantum coherence during the (non-thermal)  stringy
evaporation process,
as well as quenching
the large
Hawking-Bekenstein
entropy associated with the black hole. Defining the
latter as the measure of the loss of information
for an observer at infinity,
who - ignoring
the higher
string quantum numbers - keeps
track only of the
classical mass,angular momentum and charge of the black hole,
one recovers the familiar a
quadratic dependence on the black-hole mass by
simple counting arguments on the asymptotic density of string states
in a linear-dilaton background.}
\par
\vspace{0.2in}
\begin{flushleft}
CERN-TH.6351/91 \\
ACT-55 \\
CTP-TAMU-100/91 \\
December 1991 \\
\end{flushleft}
\end{titlepage}
\paragraph{}
\newpage
\section{Introduction and Summary}
\paragraph{}
String theory offers the possibility of resolving
once and
for all many of the deepest problems in quantum gravity,
such as finiteness, the reconciliation of quantum mechanics with
general relativity, the vanishing of the cosmological constant,
the origin and flatness of the Universe, and even the satisfactory
definition of the gravitational path integral. Varying amounts of
progress have been made in presenting the actual solutions to these
fundamental problems.
For example, it has been shown that all n-loop string amplitudes
calculated in a fixed flat background are finite \cite{mand}, but
the enumeration
of
classical non-perturbative backgrounds and understanding
of the string path integral are far from complete.
Some of these fundamental problems have been addressed from the point
of view of the effective field theory of light string states, such as
the existence of hair that might help us understand whether string
black holes respect quantum coherence \cite{hair}, and
the discovery of ``no-scale''
models \cite{nscal}
that ensure the vanishing of the cosmological constant
in some approximation. However, the resolutions of all these problems
presumably require
non-perturbative string theory techniques.
\paragraph{}
Such techniques have recently been succesfully developed and applied
to two-dimensional string quantum gravity, leading first to the
non-perturbative solution of matrix models \cite{matrix}
and more recently
to the construction of two-dimensional string black holes
\cite{gupt,witt,verl}.
Earlier constructions
exact cosmological
solutions of subcritical string theory as conformal Wess-Zumino
models had also been known \cite{aben}.
One
could hope that these non-perturbative techniques
have advanced  sufficiently for the resolutions of some of the
above-mentioned fundamental problems to be within reach. Indeed, we have
argued in a recent series of papers \cite{emn1,emn2,emn3}
that string theory reconciles
quantum mechanics and two-dimensional gravity. We have identified
an infinite set of exactly-conserved gauge quantum numbers,
``W-hair'', which are associated with the topological solitons
\footnote{These discrete states
were
discovered in the context
of matrix models by Gross and Klebanov \cite{kle}, and in the continuum
Liouville theory by Polyakov \cite{pol}. However their physical
significance, especially for black hole physics, seems not to
have
been recognized by these authors.} that form the final
stages of two-dimensional black hole evaporation \cite{emn1,gupt,witt}.
As a result of this W-symmetry, the two-dimensional
phase space volume of the matrix model
is conserved under time evolution \cite{witt2,emn2},
excluding the
introduction of modifications of the conventional
S-matrix or Hamiltonian evolution
of the density matrix \cite{ehns}
in two dimensions.
Furthermore,
we have demonstrated
explicitly that the evaporation of a two-dimensional
black hole is a
purely quantum-mechanical higher-genus
effect that does not introduce mixed states \cite{emn3}.
\paragraph{}
In this paper we extend these arguments
to argue that four-dimensional string black holes
do not lead to mixed states, and hence that string
reconciles quantum mechanics with general relativity
also in four dimensions. The central problem
of quantum coherence was seen originally for
spherically-symmetric
four-dimensional black holes
\cite{hawk2,hawk}, and the study of rotating four-dimensional
black holes has not altered the dilemma, so we emphasize
here
spherically-symmetric
four-dimensional black holes \footnote{The extension
of our arguments to rotating black holes is technically
more complicated, but presumably does not raise any
fundamental issues of principle, since string
has infinitely many gauge symmetries. See the discussion
of the Kerr solution in section 4.}. These are described
by dimensional reduction of four-dimensional
string
theories. The dimensionally-reduced
theory  is expressible as
a
$\frac{SU(1,1)}{U(1)}$ coset conformal field theory
of the two dimensional string black hole,
and thus is an
{\it exact} solution.
The radially-dependent dilaton field
enters just as in the two-dimensional model, whose
massless ``tachyon'' represents model-dependent matter fields
in four dimensions.
There is infinite-dimensional
W-hair
associated
with this $\frac{SU(1,1)}{U(1)}$
coset model\footnote{The one-dimensional axion ``hair''
found in the effective field theory approach \cite{krauss}
is just a single
strand of this infinite-dimensional W-hair, and cannot by itself
reconcile black hole dynamics with quantum mechanics.},
that conserves the 2-dimensional radial
$s$-wave phase space volume element,
and thereby prevents the appearance of non-quantum-mechanical
terms in the S-matrix description of $s$-wave scattering.
Just as in two dimensions, the quantum-mechanical
evaporation of the four-dimensional black hole
is a higher-genus effect that does not involve mixed states.
Nevertheless, by simple counting arguments on the multiplicity
of string states, one recovers the Hawking-Bekenstein formula
\cite{hawk,bek} for the apparent entropy of a black hole
($S \propto M^2 $) if one restricts oneself to the classical
conserved charges of a black hole
(mass, angular momentum
and electric
charge) and disregards the infinite string set of conserved quantities.
\section{Non-local stringy gauge symmetries}
\paragraph{}
The central obstruction to reconciling
quantum mechanics with general relativity, and thereby
avoiding the evolution of pure states into mixed states,
is the apparent
loss of information
across the horizon surrounding, e.g., a conventional
black hole.
This can be expressed mathematically via the formula \cite{hawk,bek}
\begin{equation}
      S=\frac{1}{4}k_{B}A
\label{entrop}
\end{equation}

\noindent for the entropy $S$ associated with a horizon
of area $A$ ($k_{B}$ denotes Boltzmann's constant). Using
the usual relation
between
the mass and horizon area of a spherically-symmetric
black hole, we find
\begin{equation}
     S=\frac{1}{\hbar} k_{B} M^2
\label{massentrop}
\end{equation}

\noindent The entropy (\ref{entrop},\ref{massentrop})
is just one aspect of the black hole thermodynamics
induced by quantum effects in local field theories.
There is also an apparent
temperature
\begin{equation}
T=\frac{\hbar}{8\pi M}
\label{temp}
\end{equation}

\noindent associated with a spherically-symmetric
black hole. There is an alternative statistical
definition of the entropy of a black hole :
\begin{equation}
S=-k_{B}ln N_{H}
\label{stat}
\end{equation}

\noindent where $N_{H}$ denotes the total number of
quantum-mechanical distinct ways that a
black hole,
of {\it given}
mass,
angular momentum, and charge
could have been made.
The number $N_{H}$ can be viewed as counting the number
of possible independent microscopic states of the black hole
atmosphere.
The problem is that in any local field theory there is only a
{\it finite} number of conserved gauge quantum numbers, so the
entropy (\ref{entrop}) or (\ref{massentrop}) can only be accommodated
by a mixed state. Also it is clear that a thermal state is necessarily
mixed. Recently attacks have been made on this problem
using effective field theories derived from the string,
which contain a finite number of additional conserved quantum
numbers, associated, e.g., with the axion \cite{krauss} or
with discrete gauge symmetries \cite{hair}.
However,
these still do not touch the core of the problem presented
by the entropy (\ref{entrop}) or (\ref{massentrop}), which seems
to require an infinite number of exactly conserved
quantities, if the black hole state is not to be mixed.
\paragraph{}
However, a
stringy black hole has an infinite set
of hair
associated
with the infinity of gauge symmetries that characterise
any string theory \cite{ven,kub}.
In this case
the
large entropy (\ref{entrop}) defined by Hawking
is avoided by
the following argument \footnote{Notice that
a local field theory has necessarily a finite number of
conserved charges, hence the thermal evaporation scenario
for the black holes seems the only consistent one, with all
the inevitable consequences on the loss of quantum-mechanical
coherence.}. Classically,
mass,
angular momentum and
charge
are the only type
of observable hair that a black hole can have, and hence
the necessity of a mixed state to account for the
large
entropy. In
string theories
the entropy is zero, since quantum mechanics
is valid and pure states never mix, due to
the arguments in \cite{emn1}, and
the evaporation scenario of \cite{emn3}
is {\it in vacuo}
and does not involve mixed states.
It is the information carried
by the infinite string hair
that makes the difference
from the previous calculation of
the entropy (\ref{entrop},\ref{massentrop}).
To get the latter, one
considers
only the classical charges of the black hole
and treats the (infinity of) string
gauge
charges (associated with the
the rest of the
excited string states-which in the effective two-dimensional
case are topological)  as unobservable, using them just
to count the number of
quantum-mechanically distinct ways that a black hole of given
mass, electric charge, and angular momentum is made.
In this way, eq. (\ref{stat}) accounts for the
large Hawking-Bekenstein
entropy, as we shall show explicitly in section 3.
\paragraph{}
It is instructive to
review briefly  at this point
target-space gauge
symmetries in critical
strings.
The first approach to such symmetries
was that of refs. \cite{ven,kub}, who showed that there exists
in string
theory an infinite set of generalized Ward identities
inter-relating states of different spin and mass. The lowest
such identity is
that expressing general coordinate invariance:
\begin{equation}
q^{\mu} <V^{G}_{\mu\nu}(q) \Pi_{i=1}^{N} V^{T}(k_i)>
=\sum_{i=1}^{N} k_{i\nu}
<V^T(k_i+q)\Pi_{j \ne i} V^T(k_j)>
\label{gencoor}
\end{equation}

\noindent and another involves
states of rank four, three, and two
\cite{kub}
\begin{equation}
   k^{\mu} A_{(\mu |\nu |\rho)\sigma} -iB_{\nu\rho\sigma} =
 \sum_{i=1}^{N} k_{i\nu} G_{\rho\sigma}(k_i+k, \{ k_j ; j \ne i \} )
+ perms (\nu , \rho , \sigma)
\label{rank}
\end{equation}

\noindent where $\mu|\nu|\rho \equiv \mu\nu\rho + \rho\nu\mu $.
In on-shell cases the sums on the rhs of the
above equations (\ref{gencoor},\ref{rank})
vanish on the basis of the cancelled propagator
argument \cite{green}.For our purposes
we shall only
deal
with on-shell modes, in which
case one avoids the usual ambiguities of extending
these identities off string-shell \cite{kub}.
Ref. \cite{ovr} gave a conformal field theory analysis
of such gauge symmetries, showing that there was one
associated with every $(1,0)$ or $(0,1)$ operator, of which string
theories have an infinite number. Contained within this infinite set
of gauge symmetries is the particular $W_{\infty +1}$ symmetry
located in studies of two-dimensional string gravity, whose associated
``ground ring'' algebraic structure of $(1,0)$ and $(0,1)$ operators
has been discussed in ref. \cite{witt2}.
\paragraph{}
A simple counting argument indicates that the number
of such $(1,0)$ and  $(0,1)$ operators in a four-dimensional
string theory is comparable to the entropy (\ref{massentrop})
of a massive black hole, and hence might be adequate to accommodate
this entropy without the necessity of a mixed state.
This is based on the fact that the number of gauge
symmetries is
at least in correspondence
with the number of string levels.
For example, a subclass
of
$(1,0)$ operators corresponding to string
level $2N$ assumes the generic form
\cite{ovr}
\begin{equation}
  \int d\sigma \Psi (\partial X)^{N} (\partial X)^{N-1}
\label{index}
\end{equation}

\noindent where $\sigma$ is a space-like world sheet parameter,
and $\Psi$ is a $(2N-1)$-index tensor space-time field
that is symmetric on the first $N$ and last $N-1$ indices,
as well as divergence-free on each index. Clearly this
is an infinite set of operators in any string theory.
The actual symmetries are bigger \cite{ovr}. On the
basis of general arguments,
one
could expect that each gauge stringy symmetry would lead
to a conserved charge, which could participate
in characterising the black hole.
\paragraph{}
In two dimensions, the existence of an infinite set
of conserved quantum numbers was first demonstrated
in matrix models \cite{kle}. We subsequently pointed
out that they should also appear as gauge symmetries
of two-dimensional black holes \cite{emn1}, associated with the
massive {\it topological} discrete discrete states that are known to
exist in continuum Liouville models of two-dimensional gravity
\cite{pol}. Indeed, the gauge nature of these symmetries
was subsequently demonstrated in ref. \cite{seib}.
We argued \cite{emn1} that this infinite set of conserved
charges, ``W-hair'', should be sufficient to maintain
quantum coherence for two-dimensional black holes,
consistently with the known
existence of an $S$-matrix
for two-dimensional matrix models. This quantum-mechanical
behaviour was subsequently given an elegant geometrical
interpretation in terms of an infinite phase-space area-preserving
symmetry \cite{emn2}, which
originates from the ground ring
of $(1,0)$ or $(0,1)$ world-sheet operators mentioned earlier
\cite{witt2}. In confirmation of this point,
it was shown subsequently \cite{emn3}
that the quantum evaporation of a two-dimensional
black hole was related to the imaginary part of a formally
divergent higher-genus string amplitude, associated with an integral
over large
tori.This evaporation did not have a finite-temperature
interpetation, and did not lead to a mixed state.
For the purposes of the later discussion we note that
this evaporation mechanism did not seem specific
to two dimensions, and appeared to be generalizable
to four-dimensional black holes.

\section{Spherically-Symmetric Four-dimensional Black Holes}
\paragraph{}
We note first that the original arguments given by Hawking
referred to spherically-symmetric black holes originated by
the spherically-symmetric collapse of macroscopic matter
\cite{hawk}.
Spherically-symmetric solutions to gravity theories
in arbitrary dimensions have been classified in a wide class
of theories \cite{whitt}. In particular, the so-called
second-order formalism has been adopted for a description
of gravity theories in arbitrary number of dimensions,
involving
in general
higher powers of the curvature tensor.
The result is that, with the exception of some
unphysical
cases, all spherically-symmetric
solutions are {\it static} \cite{whitt}
and some of them are known to exhibit
singularities hidden by event horizons, and therefore are of
black hole type. Since all such spherically-symmetric
singularities can be regarded as in some sense
two-dimensional, the angular variables
being inessential, we analyze them using results from
string theory in two-dimensional space-time, which we now review
briefly.
\paragraph{}
Witten \cite{witt}
showed that it is
possible
to describe the region of two-dimensional target
space-time around the singularity by an {\it exact}
conformal field theory, which is a
coset $\frac{SU(1,1)}{U(1)}$
Wess-Zumino
$\sigma$-model formulated on an arbitrary Riemann surface $\Sigma$.
In \cite{emn3} we have argued that summation over
Riemann surfaces of arbitrary topology, as required by a consistent
string formalism, produces modular infinities which have to be
regularised by analytic continuation, thereby leading to
imaginary mass shifts of the black hole solution and hence
{\it instabilities}. The latter will cause the black hole to evaporate,
but such an evaporation, although quantum in origin, is different
from the thermal scenario argued by Hawking \cite{hawk}
in conventional {\it local} field theories of gravity.
The evaporation is necessitated by the fact that the
string black hole solutions carry an infinite
number of {\it conserved} quantum numbers, arising
from stringy {\it gauge}
symmetries \cite{ven,ovr,emn1} mixing the various mass
levels. In the
case of two-dimensional
black holes these charges are known \cite{seib,emn2}
to form a
$W_{\infty+1}$ extended conformal algebra.
This symmetry is a subgroup of an area-preserving
infinite dimensional algebra generated by
{\it world-sheet} currents of conformal spin $(1,0)$ or $(0,1)$
\cite{witt2}. Due to this fact, we have shown in \cite{emn2}
that the symmetry is elevated into a target space one
leading to the infinity of conserved
charges mentioned before. The reason is that such world-sheet
symmetries constitute a
{\it canonical} deformation
of the
conformal field theory ( stringy $\sigma$-model)
describing the world-sheet dynamics \cite{ovr}.
The latter are
represented as induced transformations
of the (target space) background fields of the $\sigma$-model.
The important feature, relevant for issues of
quantum coherence, is that this symmetry preserves
the phase-space area of the matrix model \cite{witt2},
which
describes
the interaction
of $c=1$ matter
with the black hole. In \cite{emn2}
we pointed out that it is precisely this property of the
infinite-dimensional string symmetry that ensures preservation of
quantum
coherence during the black hole evaporation process. This was
the
feature
that was believed to be violated
according to the
Hawking arguments \cite{hawk2}
on the non-factorisability of the
conventional scattering matrix due to the presence of
space-time
singularities \footnote{According to
Hawking \cite{hawk2} these constituted
an obstruction to the analytic continuation
from Euclidean to Minkowskian space, causing
the non-factorisability property. Such modifications
would invalidate the $CPT$-theorem
of quantum mechanics in its ordinary sense,
as the later is
incompatible with a non-factorisable
\$ - matrix. However if one abandons the
concept of  a superscattering operator,
while keeping the density matrix
formalism as fundamental, then $CPT$-invariance
can be preserved \cite{page}, perhaps at the
cost
of not having definite mixed states, a situation
even less deterministic than that of Hawking \cite{hawk}.
In our case, the factorisability of the Hawking
matrix \$ is guaranteed due to symmetries of the theory
and hence $CPT$-invariance holds in the strong (ordinary)
sense.}
\paragraph{}
These arguments were originally formulated in
two-dimensional
string cases, and one can naively think that they have
nothing to do with the real four-dimensional case. However we shall
now argue
that this is not the case,
since spherically-symmetric
solutions of four-dimensional gravity theories
are effectively {\it two-dimensional} theories.
We conjecture that, in order to describe the
spherically-symmetric singularities one can use the formalism of
two-dimensional strings. This conjecture seems also to
be in agreement with Witten's point of view \cite{witt3}.
However, as we argued  already in \cite{emn1,emn2}
and
we shall repeat below, it seems to us that full
consistency of general relativity with quantum
mechanics is achieved only upon inclusion of the
{\it entire} spectrum of topological string states,
which in two dimensions constitute the remnants
of excited string states in higher-dimensional
target spaces.
\paragraph{}
To be systematic, we start from the observation
that in a  $D$-dimensional target-space
string theory there is an infinity of
{\it discrete} topological states, which are
similar in nature to
those of the two-dimensional case
\cite{pol}. Indeed these states can be seen in
the {\it gauge} conditions for a rank $n$ tensor multiplet,
\begin{equation}
   D^{\mu_{1}}A_{\mu_{1}\mu_2...\mu_n}=0
\label{gauge}
\end{equation}

\noindent  where $D_{\mu}$ is a (curved space) covariant derivative.
To illustrate our arguments,
consider the simplified case of
weak gravitational perturbations around
flat space, with a linear dilaton field
of the form $\Phi(X)=Q_{\mu}X^{\mu}$. One finds the following
Fourier transform of (\ref{gauge}),
\begin{equation}
    (p + Q)^{\mu_1} {\tilde A}(k)_{\mu_1\mu_2....\mu_n}=0
\label{gauge2}
\end{equation}

\noindent We then observe that there is a jump in the number
of degrees of freedom at discrete momenta $p=-Q$.
Due to the
complete uncertainty in space, such states are delocalised,
and can be considered as quasi-topological and
non-propagating
soliton-like states. In ordinary string theories,
such states
presumably
carry a
small statistical weight, due to the
continuous
spectrum of the various string modes.
However,
in strings propagating in
spherically-symmetric four-dimensional
background space-times,
these discrete states
become {\it relevant}. Such backgrounds are effectively
two-dimensional,
and therefore all the
transverse modes of higher rank tensors can be gauged away
using
Ward identities of the form (\ref{gauge}), except for
the {\it topological} modes. In a four-dimensional
spherically-symmetric background formalism, these are $s$-wave
topological modes. For spherically-symmetric black holes,
these
modes constitute the final stage
of the evaporation \cite{emn1,gupt},
and they are
{\it responsible} for the maintenance of
quantum coherence \cite{emn1,emn2}.
For clarity we shall recapitulate the arguments
of \cite{emn1,emn2,emn3}, emphasizing
that now one is really dealing with
four-dimensional
space-time spherically-symmetric singularities.
\paragraph{}
The analysis of \cite{whitt}
implies that in pure gravity all the {\it classical}
spherically-symmetric solutions to the equations of
motion obtained from
higher-derivative
gravitational actions with an arbitrary
number of curvature tensors
are {\it static}.
A similar result occurs in the case of
string-theoretic
black holes at tree string-level. However,
the arguments of \cite{emn3}
imply instabilities
in spherically-symmetric black hole solutions, since
these are massive string states and as such should
be able
to decay to lighter states \cite{marc,turok}.
This
mechanism
also exists
for superstring theories.
Consider then a superstring
theory, and a spherically-symmetric gravitational
background
of black hole type.
The metric tensor will be given by an Ansatz of the form:
\begin{equation}
ds^2 =g_{\alpha\beta} dx^{\alpha}dx^{\beta} + e^{W(r,t)}d\Omega^2
\label{sphere}
\end{equation}

\noindent where W(r,t) is a non-singular function
and $x^\alpha,  x^\beta$ denote $r,t$
 coordinates. Also,
 $d\Omega ^2 = d\theta ^2 + sin^2 \theta d \phi ^2 $
 denotes the line element
on a spherical surface that does
not change with time. It can be shown that the standard
Schwarzschild solution of the spherically-symmetric four-dimensional
black hole \cite{thorn} can be put in
the above form
by
an appropriate transformation of variables.
Consider the Schwarzschild solution in Kruskal-Szekeres
coordinates \cite{thorn}
\begin{equation}
ds^2 = -\frac{32M^3}{r}e^{-\frac{r}{2M}}du dv + r^2 d\Omega ^2
\label{krusk}
\end{equation}

\noindent
Here $r$ is
a function of $u,v,$ since it is given by
\begin{equation}
  (\frac{r}{2M} - 1) e^{\frac{r}{2M}}=-uv
\label{reln}
\end{equation}

\noindent Notice that despite the static character
of the
black hole solution, upon changing variables,
the two-dimensional metric components depend
on {\it both} variables $u,v$.
\par
\noindent Changing variables to
\begin{eqnarray}
\nonumber
e^{-\frac{r}{4M}}u=u' \\
e^{-\frac{r}{4M}}v=v'
\label{change}
\end{eqnarray}

\noindent and
taking into account the Jacobian $J$
of the transformation
in the (positive-definite) area element $dudv$, we
can put
the two-dimensional metric in the form
\begin{equation}
 g_{bh}(u',v')= \frac{e^{D(u',v')} du'dv'}{1-u'v'}
\label{bh}
\end{equation}

\noindent
where the scale factor
is given by $ 16M^2 e^{-\frac{r'(u',v')}{2M}}J(u',v')$,
with $r'$
the function $r$ expressed in $u',v'$ coordinates.
This form of the metric is just a conformally-rescaled
form of Witten's two-dimensional  black hole solution
\cite{witt}. Since the
latter is described by an
exact conformal field theory, so is the conformally-rescaled
metric, which from a $\sigma$-model point of view simply
expresses a sort of renormalisation scheme
change\footnote{The function $D(u,v)$ can be regarded also as
a part of the two-dimensional dilaton in the given renormalisation
scheme.}. From now
on we shall work directly with the
conformally-rescaled metric.
The global properties (singularities) remain unchanged
from
the two-dimensional string case. In particular, according to
the interpretation of
Witten's work \cite{witt} by Eguchi \cite{eguchi},
a
(conformal)
Wess-Zumino
coset
model is suitable for the description of the
region of target space-time around the singularity, where
the conventional $\sigma$-model formalism breaks down.
\paragraph{}
To understand this point better, let us consider
the gravitational sector of
a four-dimensional supersymmetric string effective action
It
has the generic form
\begin{equation}
  S_{eff}=\int d^4x \sqrt{G} e^{\Phi} \{ \frac{1}{\kappa ^2}
  R^{(4)} - \frac{1}{2} (\nabla_{\mu}
   \phi)^2 - e^{-2\sqrt{2}\kappa \phi} H_{\mu\nu\rho}^2 + ... \}
\label{sugra}
\end{equation}

\noindent where $\phi$ is a four-dimensional dilaton field,and
$H_{\mu\nu\rho}=\partial_{[\mu}B_{\nu\rho]} + \omega_L -
\omega_Y $  is the field
strength
of an antisymmetric tensor field, which by a duality
transformation, upon using the
equations of motion,
is equivalent to a pseudoscalar $\lambda$.
The dots ... denote higher-derivative terms as well as gauge
or other matter fields coming from compactification,
in the case that
one starts from a
string theory in the critical dimension.
Their presence does not affect our discussion.
Upon dimensionally reducing (\ref{sugra}) to
discuss the spherically-symmetric gravitational background,
one observes that {\it another} dilaton ($W(r,t)$)
is going to
be generated by the angular part of the Ansatz
(\ref{sphere}), as well as a two-dimensional
{\it cosmological} constant term, even if
the
four-dimensional theory has zero cosmological constant.
The reasons are simple. Since the metric is spherically-symmetric,
there will be the radial part
$g_{\alpha\beta} $ in (\ref{sphere})
(depending on time in general),
which
yields
a two-dimensional
scalar curvature term $R^{(2)}$.
{}From the four-dimensional
determinant one obtains
exponential $W$-dilaton factors accompanying the two-dimensional
metric determinant $\sqrt{g}$, and from the
derivatives with respect
to $r$ and $t$ of the angular part
one gets two-dimensional $W$-dilaton kinetic terms.
The
angular part of the metric (\ref{sphere}),
with
constant two-dimensional curvature,
yields a
cosmological constant part\footnote{
The $d\Omega ^2$ represents the metric of a 2-sphere
of unit radius, with scalar curvature $2$.
Integration over the angular variables yields
an extra factor of $4\pi$.}.
Thus the
effective description of the theory is given by
the following
two-dimensional
effective action (to lowest order in derivatives)
\begin{equation}
4\pi
\int d^2 x e^{W} \sqrt{g} (R^{(2)} - (\nabla W)^2 - 2
                                     + (\nabla T)^{2} +...)
\label{effective}
\end{equation}

\noindent We are interested in the extra charges that the
black-hole  can have in a string effective model.
In the case of spherical geometry,
this implies a
spherically-symmetric
Ansatz for the matter fields in (\ref{sugra}).
This leads to scalar $s$-mode structures for the antisymmetric
tensor (axion) and higher-dimensional dilaton or other
matter fields' $s$-wave modes which are collectively represented as a
two-dimensional string ``tachyonic'' mode, $T$
\footnote{We should stress that in our formalism
the ``dilaton'' of the
two-dimensional string model
occurs necessarily in the spherical Ansatz for
four-dimensional
gravity,
and therefore one does not have to start from a
higher-dimensional string model
and compactify. All such compactification
modes that occur in
traditional superstring-inspired models
\cite{witten}
appear in our two-dimensional effective model
as matter ``tachyon'' $T$-fields.}.
\paragraph{}
Having expressed the theory as a two-dimensional effective string model,
one can apply the whole machinery of two-dimensional strings
to study the dynamics of the evaporation of the black holes
and determine the final stage. It should be stressed
that,
from the general analysis mentioned in the
beginning \cite{whitt},
the static character of the
physically-interesting solutions to string-inspired
gravitational theories implies probably that
the only way that these black holes evaporate
is the one suggested in \cite{emn3}, i.e. through string
quantum corrections, requiring a
formulation in higher world-sheet
genera and summation over them. Independent
arguments to support this claim
will be given below. At present,
we note that
such decay is non-thermal,
and hence maintains
quantum coherence,
as guaranteed by
the
$W_{\infty}$-symmetry
associated with the discrete topological $s$-wave modes.
The $s$-wave matter phase-space,
which would be the problematic
one from the point of view of quantum coherence, due to
modes going into and not coming out or vice versa,
is two-dimensional in the spherically-symmetric case
and hence the arguments of ref. \cite{emn2}
apply. The symmetry of the effective
two-dimensional
target space is phase-space volume (area
in two dimensions)
preserving, and hence Liouville's theorem for the time
evolution of the density matrix remains valid.
Thus, there is no modification of the evolution
equation of the density matrix in the presence of a spherically
symmetric black hole \cite{ehns}, and the
factorisation of Hawking's superscattering
operator holds.
\paragraph{}
A further comment concerning the thermodynamical relations
(\ref{entrop}), (\ref{massentrop})
of the black hole solutions is in order.
We noted in \cite{emn3} that the quantum instabilities
associated with modular infinities, which cause the
evaporation of two-dimensional
black holes
(and hence of spherically-symmetric
four-dimensional configurations as well), are non-thermal
in origin. Arguments have been given \cite{emn3}
for the thermal
stability
of these objects on the basis of compactified $c=1$ matrix
models, believed to represent a
stringy regularisation of two-dimensional strings (and,
in view
of the picture in this article,
of four-dimensional strings propagating
in spherically-symmetric backgrounds). Here we would like to give
an independent argument in support of the absence of thermal
evaporation, at least in the conventional sense,  by showing
that the available string states
account
for the quadratic mass dependence of
the black hole entropy (\ref{massentrop}).
\paragraph{}
The argument is based on the fact that
the black hole is a particular
string state of mass $M$. The Hawking
entropy is viewed as the number of ways $N(M)$
one can construct a state of this mass
(ignoring the
associated string $W$
quantum numbers, which, in view of our previous
arguments, would make the exact string entropy vanish).
In this picture $N(M)$ may be considered the same
as the multiplicity of
string states
of mass level $M$.
This entropy  is measured by an observer at spatial
infinity, where the string propagates in a flat
background with a linear dilaton field, $Q_{\mu}X^{\mu}$.
In the two-dimensional black hole case the black hole
mass is determined by
a constant shift $2a$
in the dilaton
which is non-trivial.
The precise relation is
\cite{witt,gupt}
\begin{equation}
 \sqrt{\alpha '}M=Qe^{2a}
\label{bhmass}
\end{equation}

\noindent A
choice of $a$
selects a {\it particular} black-hole
configuration (i.e. a vacuum for the
string). If we rescale the Regge
slope by $e^{-2a}$, then in units
of the rescaled slope the black hole
mass is given by $Q$.
The situation is then analogous to that of ref. \cite{aben},
where for large mass levels the multiplicities are
given asymptotically by
\footnote{We should stress that the same
arguments of \cite{aben}
apply here, despite the Wick-rotated $Q$ relative to
their case.}
\begin{equation}
   N(M)= lim_{a,M \rightarrow \infty}
   e^{2\pi \sqrt{\alpha '}
    \sqrt{2 + Q^2e^{2a}}M}=e^{2\pi \alpha 'M^2}
\label{multi}
\end{equation}

\noindent where $\alpha '$ is the redefined Regge slope,
depending on the particular black hole background.
There is no loss of generality if we express the black hole mass
in units of this.
The entropy is
determined
by taking the logarithm $-k_BlnN(M)$, thereby leading
to quadratic mass dependence of the
form (\ref{massentrop}),
as argued by Hawking and Bekenstein
\cite{hawk,bek}
using classical thermodynamics
and information theory. It should be noticed that
as the mass decreases there will be a point where
formula (\ref{multi}) will be no longer applicable,
and thus the situation is analogous to that of Hawking
\cite{hawk} where quantum effects become important for small
black hole
masses and classical thermodynamics arguments cannot be used.
\paragraph{}
So far we have dealt with static black holes.
Let us now consider the above thermodynamical
relations in connection with
the evaporation mechanism
described in \cite{emn3}. The latter is
based on the fact that
{\it any} massive string state {\it decays} to lighter
states in such a way that the stringy gauge symmetries
associated with the relevant mass levels remain intact.
The formal origin of such an instability appears as a modular
divergence of the two-point function of the state in question.
Regularisation of the infinity by means of, say, analytic
continuation yields an imaginary part $I$ \cite{marc,turok}. The
latter, in view of the
validity of the $S$-matrix  formalism due to the $W$-hair property
of the string black holes, implies due to the optical theorem
a decay with a decay rate $\Gamma=-2I$.
The
dimensionality of space-time plays a crucial
r\^ ole in determining the life-time. To see this,
let us consider
a simple field theory example, which however captures
all the essential features of the string case. Consider
\cite{turok} two fields $\Phi$ and $\phi$
of masses $M$ and $m$ respectively, with
$M>2m$, and a ``string inspired''
$\frac{1}{2}\lambda \Phi \phi^2 $ interaction.
The imaginary part of the one-loop two-point function
of the state $M$ can be computed in terms of $M,m$ \cite{turok}
\begin{equation}
     \Gamma \equiv \frac{1}{M}\frac{dM}{dt} =
      \frac{\lambda ^2 \pi  M^{D-5}}{ (16 \pi)^{\frac{D-1}
{2}} \Gamma(\frac{D-1}{2})} (1-\frac{4m^2}{M^2})^{\frac
{D-3}{2}}
\label{decay}
\end{equation}

\noindent The above computation necessarily
goes off-mass shell for the propagating
light particle in the loop.
In the two-dimensional case one observes
that the decay rate $\frac{dM}{dt}$
is proportional to $M^{-2}$. This property,
when transcribed
into the case of black hole decay {\it in vacuo} yields
the entropy relation (\ref{massentrop}), with coefficient
dependent on the details of the process
\footnote{The proportionality
coefficient depends on
detailed string
computations, to which we hope to return
in the near future.}.
In string theory one has an infinity of states propagating
in the loop. However,
in the two-dimensional {\it effective
string theory} the only propagating states are those of the
massless ``tachyons'' \cite{pol}. The rest of the
string states are
topological. As a crude estimate therefore of the corresponding
decay rate we take (\ref{decay}) with $m=0$, $d=2$ and
$M$ the black hole mass. Although detailed computations
must be done to estimate the magnitude of the proportionality
coefficients,
we believe the heuristic argument
we gave above
is sufficient to demonstrate the essential
physics of black hole evaporation.
Although the evaporation is not thermal,
one can give a thermodynamic interpretation
\`a la Hawking \cite{hawk}, if
one restricts oneself to the entropy
defined in the classical black hole case (see the
discussion
in section 2 and in the previous paragraph). Then,
the thermodynamic relation (\ref{massentrop}) follows
from the
simple fact that the decay {\it in vacuo}
occurs with a rate proportional to $M^{-2}$.
Upon time-averaging during the decay from an initial
black hole mass $M$ down to
mass zero, we observe that
the average Hawking entropy still obeys a
quadratic-mass law. This is a feature only
of the inverse squared-mass behaviour of the
decay rate.
As a side remark,
we would like to
point out that in local field theories the relation
(\ref{massentrop}) can be also
explained
in a thermal
way by considering the evaporation of the black hole
{\it in equilibrium} with a heat bath of temperature
approaching the Hawking one from below \cite{york}.
In such a scenario,
the crucial point for getting a
large statistical entropy
is the zero-point
energy of {\it summed up}
excited states
which disappear if we allow the black
hole
to evaporate down to a final mass zero.
This is to be compared with
the situation in the present case,
where however the evaporation/decay
takes place {\it in vacuo}.
The topological states, which are responsible for the
maintenance of quantum coherence during the evaporation
process, also ``disappear''' as external states
in the limiting zero-mass case
represented by
matrix model.
However,
their presence is essential in yielding
the enormous statistical entropy for the stringy black hole.
In the formal sense, their presence
guarantees
the correct value of the proportionality
coefficient
in (\ref{massentrop}) to match
in order of magnitude that of Hawking's original
computation, based on classical black body radiation \cite{hawk}.
\paragraph{}
\section{Non-Spherically-Symmetric Black Holes}
What about spherically non-symmetric
singularities ?
Such objects are known as solutions of
Einstein's equations, the Kerr rotating black holes
for example \footnote{For classical Einstein gravity
coupled to Maxwell's electromagnetism
there are uniqueness theorems for rotating black holes
\cite{mazur}. The situation is less clear for
quantum theories, especially the ones obtained as a low energy
limit of string theories, where higher-derivative modifications
of Einstein-Maxwell's equations occur. For our purposes we shall
concentrate on the Kerr
solution, which is the only one
studied extensively so far. We believe that this is
sufficient to demonstrate our arguments.}. In
the context of the present formalism,
rotating objects with event horizons
can be constructed by appropriate
tensoring of Wess-Zumino models in two dimensions \cite{hor}.
The original papers
on the possibility of the loss of quantum coherence have
concentrated on the spherically-symmetric case, which we have
argued does not have any such
problems in the string case.
What happens in spherically-non-symmetric cases is not yet
fully understood. However, we now
note that even in the case of Kerr
black holes, one can argue that in
physically-interesting
cases the final stage of the evaporation excites
$s$-wave
topological modes that are similar to the
spherically-symmetric case.
\paragraph{}
Consider the Kerr
metric \cite{stef}
\begin{eqnarray}
\nonumber
ds_{Kerr}^2 = \frac{r^2 + A^2 cos^2 \theta - 2Mr}{r^2 + A^2 cos^2 \theta}
dt^2 -
\frac{r^2 + A^2 cos^2 \theta}{r^2 + A^2 - 2Mr} dr^2
 - (r^2 + A^2cos^2 \theta ) d\theta^2 - \\
\frac{[(r^2 +  A^2 )^2 - A^2 sin^2 \theta (r^2 + A^2 -2Mr)]sin^2 \theta}
{r^2 + A^2 cos^2\theta} d\phi^2
- \frac{4MA r sin^2 \theta}{r^2 + A^2 cos^2\theta} dt d\phi
\label{kerr}
\end{eqnarray}

\noindent where $M$ is the mass of the black hole, and
$A=\alpha M$ is the angular momentum. For physically-interesting
cases \cite{stef,mazur} $ 0 < \alpha^2 < M^2 $.
The region $M^2 < \alpha^2 $
corresponds
to very rapid
rotation of the
body, which probably does not occur for
real physical bodies,
as they would fly apart
before rotating so rapidly. The final stage of the evaporation,
$M \rightarrow 0$, therefore, would correspond to $\alpha \rightarrow
0$ as well. Expanding in powers of $M$ and keeping only the
leading order it is straightforward to see from (\ref{kerr})
that the limit is again
one with a topological $Q$-graviton
as in the two-dimensional string case (upon redefining
$r \rightarrow e^Q\rho$, where $Q$ is $2\sqrt{2}$, as required by the
conformal field theory interpretation of the spherically-symmetric case
\cite{gupt,emn1}). Perturbations around the Kerr solution by the
other stringy modes will presumably lead to an
excitation of the rest of the topological $s$-wave modes
in the final stage of the evaporation, which is similar
to that in the
spherically-symmetric case.
\paragraph{}
This oversimplified argument
suggests
that $W$-symmetry,
or rather a generalised
form of it, appropriate for spherically-non-symmetric matter,
generated by topological (discrete)
stringy modes,
characterises
in general
singularities in four-dimensional target space-times.
The formal reason is that the latter are described by
topological theories which are
in some sense
characterised by an
infinite-dimensional
extended conformal symmetry. We expect that the
generalisation (to non-symmetric spaces)
of the
area-preserving $W$-symmetry
characterising the two-dimensional symmetric case
will be
a phase-space volume preserving algebra. In the same way that
W-symmetry characterises $\frac{SU(1,1)}{U(1)}$ coset
Wess-Zumino models \cite{bakas}, one might find that
the appropriate extension of these theories to describe
spherically-non-symmetric matter would be
phase-space volume-preserving groups. This,
however, is
still a speculation, but one expects
that the quantum coherence
problem can be solved in general due to the enormous
stringy symmetries,
without reference to any
particular geometry for the singularity.

\section{Conclusions}
\paragraph{}
We have argued that spherically-symmetric four-dimensional
objects with physical curvature singularities and event horizons
(black  holes) can be described by effective two-dimensional
string theories. Such models have necessarily
a dilaton field, and can be represented as coset
$\frac{SU(1,1)}{U(1)} $ Wess-Zumino models, which are known to
possess
$W$-symmetries that preserve the
two-dimensional $s$-wave
phase-space
under time evolution, and thus maintain
quantum coherence during the (non-thermal) stringy evaporation
process of a spherically-symmetric four-dimensional black hole.
The latter expresses decay
of the massive stringy black hole state due to instabilities
induced by the string propagation in summed up world-sheet
topologies. We have argued that such an evaporation
is consistent with a
large Hawking entropy associated
with the black hole, the latter being viewed as
the entropy the black hole appears to have if one does not
take
into account
the infinity of (observable)
quantum charges associated with topological
excited string states.
The quadratic black-hole-mass dependence of this
entropy, conjectured
by
Bekenstein and
Hawking \cite{bek,hawk}, can be obtained
from simple formulas \cite{aben}
yielding
the
asymptotic density of string states
in a linear dilaton background, which
resembles the asymptotic form of the black-hole
space-time.
\paragraph{}
\noindent {\Large{\bf Acknowledgements}} \\
\par
One of us (J.E.) thanks the University of Miami
Physics Department for its hospitality while
this work was being completed. The
work of D.V.N. is partially supported by DOE grant
DE-FG05-91-ER-40633.

\end{document}